\documentclass[final,5p,times,twocolumn]{elsarticle}
\usepackage{amssymb}

\begin{document}

\begin{frontmatter}

\title{Intrinsic Correlation between Hardness and Elasticity in
Polycrystalline Materials and Bulk Metallic Glasses}

\cortext[cor1]{Corresponding author}
\author{Xing-Qiu Chen\corref{cor1}}
\ead{xingqiu.chen@imr.ac.cn}
\author{Haiyang Niu}
\author{Dianzhong Li\corref{cor1}}
\ead{dzli@imr.ac.cn}
\author{Yiyi Li}

\address{Shenyang National Laboratory for
Materials Science, Institute of Metal Research, Chinese Academy of
Sciences, Shenyang 110016, China}

\begin{abstract}
Though extensively studied, hardness, defined as the resistance of a
material to deformation, still remains a challenging issue for a
formal theoretical description due to its inherent mechanical
complexity. The widely applied Teter's empirical correlation between
hardness and shear modulus has been considered to be not always
valid for a large variety of materials. Here, inspired by the
classical work on Pugh's modulus ratio, we develop a theoretical
model which establishes a robust correlation between hardness and
elasticity for a wide class of materials, including bulk metallic
glasses, with results in very good agreement with experiment. The
simplified form of our model also provides an unambiguous
theoretical evidence for Teter's empirical correlation.
\end{abstract}


\end{frontmatter}

\section{Introduction}

Despite the great efforts, to understand the theory of hardness and
to design new ultrahard materials is still very challenging for
materials scientists\cite{Tse,Sup,Szymanski,Gilman2}. During the
past few years, several semi-empirical theoretical models
\cite{Gao02,Simunek,Mukhanov,Li08,Smedskjaer} have been developed to
estimate hardness of materials based on: ($i$) the bond length,
charge density, and ionicity \cite{Gao02}, ($ii$) the strength of
the chemical bonds \cite{Simunek}, ($iii$) the thermodynamical
concept of energy density per chemical bonding \cite{Mukhanov}, and
($iv$) the connection between the bond electron-holding energy and
hardness through electronegativity \cite{Li08}, and ($v$) the
temperature-dependent constraint theory for hardness of
multicomponent bulk metallic glasses (BMGs) \cite{Smedskjaer}.
Experimentally, hardness is a highly complex property since the
applied stress may be dependent on the crystallographic
orientations, the loading forces and the size of the indenters. In
addition, hardness is also characterized by the ability to resist to
both elastic and irreversible plastic deformations and can be
affected significantly by defects (\emph{i.e.}, dislocations) and
grain sizes \cite{Gilman}. Therefore, hardness is not a quantity
that can be easily determined in a well-defined absolute scale
\cite{Tse}. It has been often argued \cite{Gao01} that hardness
measurements unavoidably suffer of an error of about  10\%. All
these aspects add huge complexity to a formal theoretical definition
of hardness \cite{Gao02,Simunek,Mukhanov,Li08,Smedskjaer}.

\begin{figure}[hbt]
\begin{center}
\includegraphics[width=0.40\textwidth]{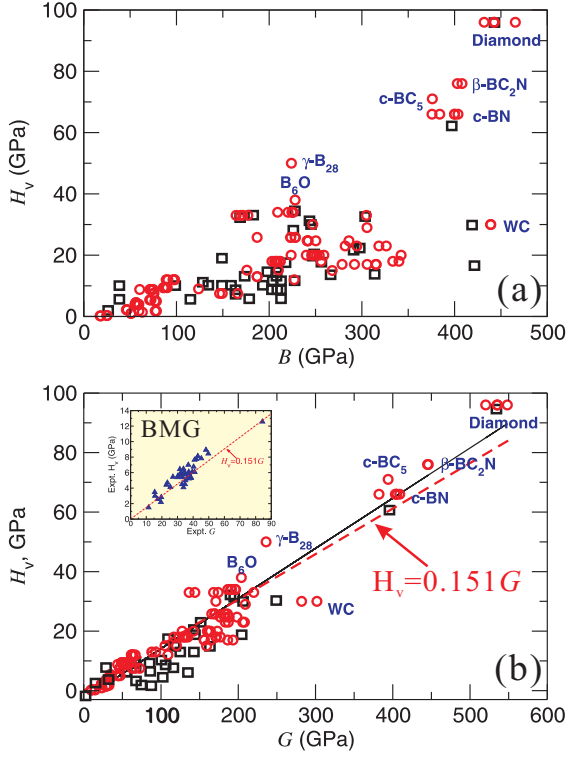}
\end{center}
\caption{(Color online) Correlation of experimental Vickers hardness
($H_v$) with (a) bulk modulus ($B$) and with (b) shear modulus ($G$)
for 39 compounds (Tables \ref{tab:2} and \ref{tab:3}). Inset of
panel (b): \emph{H}$_v$ vs. \emph{G} for 37 BMGs (see Table
\ref{tab:1}). The solid line denotes empirical Teter's fitting
values, whereas dashed lines correspond to the value derived from
Eq. (\ref{5}). The black and hollow squares denote data taken from
Refs. (\cite{Teter,Tse}).
} \label{fig1}
\end{figure}

Within this context, to find a simple way to estimate hardness of
real materials is highly desirable. Unlike hardness, the elastic
properties of materials can be measured and calculated in a highly
accurate manner. Therefore, it has been historically natural to seek
a correlation between hardness and elasticity. The early linear
correlation between the hardness and bulk modulus (\emph{B}) for
several covalent crystals (diamond, Si, Ge, GaSb, InSb) was
successfully established by Gilman and Cohen since 1950s
\cite{Gilman,Cohen}. Nevertheless, successive studies demonstrated
that an uniformed linear correlation between hardness and bulk
modulus does not really hold for a wide variety of
materials\cite{Teter,Tse,Gao01}, as illustrated in Fig.
\ref{fig1}(a). Subsequently, Teter\cite{Teter} established a better
linear correlation between hardness and shear modulus ($G$), as
illustrated in Fig. \ref{fig1}(b). This correlation suggests that
the shear modulus, the resistance to reversible deformation under
shear strain, can correctly provide an assessment of hardness for
some materials. However, this correlation is not always successful,
as discussed in Refs. \cite{Gao02, Gao01, He04}. For instance,
tungsten carbide (WC) has a very large bulk modulus (439 GPa) and
shear modulus (282 GPa) but its hardness is only 30 GPa
\cite{Haines}, clearly violating the Teter's linear correlation [see
Fig. \ref{fig1}(b)] \cite{Gao02}. Although the link between hardness
and elastic shear modulus can be arguable, it is certain to say that
the Teter's correlation grasped the key.

In this manuscript, following the spirit of Teter's empirical
correlation, we successfully established a theoretical model on the
hardness of materials through the introduction of the classic Pugh
modulus ratio of \emph{G}/\emph{B} proposed in 1954 \cite{Pugh}. We
found that the intrinsic correlation between hardness and elasticity
of materials correctly predicts Vicker's hardness for a wide variety
of crystalline materials as well as BMGs. Our results suggest that,
if a material is intrinsically brittle (such as BMGs that fail in
the elastic regime), its Vicker's hardness linearly correlates with
the shear modulus ($H_v$ = 0.151 $G$). This correlation also
provides a robust theoretical evidence for the famous empirical
correlation observed by Teter in 1998. On the other hand, our
results demonstrate that the hardness of crystalline materials can
be correlated with the product of the squared Pugh's modulus ratio
and the shear modulus (\begin{math} H_v =
2(k^2G)^{0.585}-3\end{math} where $k$ is Pugh's modulus ratio). This
formula provides the firm evidence that the hardness not only
correlates with shear modulus as observed by Teter, but also with
bulk modulus as observed by Gilman {\em et al}. Our work combines
those aspects that were previously argued strongly, and, most
importantly, is capable to correctly predict the hardness of all
compounds included in Teter's \cite{Teter}, Gilman's
\cite{Gilman2,Gilman}, Gao's \cite{Gao02} and Simunek's
\cite{Simunek} sets. Also, our model clearly demonstrates that the
hardness of bulk metallic glasses is intrinsically based on the same
fundamental theory as the crystalline materials. We believe that our
relation represents a step forward for the understanding and
predictability of hardness.

\begin{figure}[hbt]
\begin{center}
\includegraphics[width=0.40\textwidth]{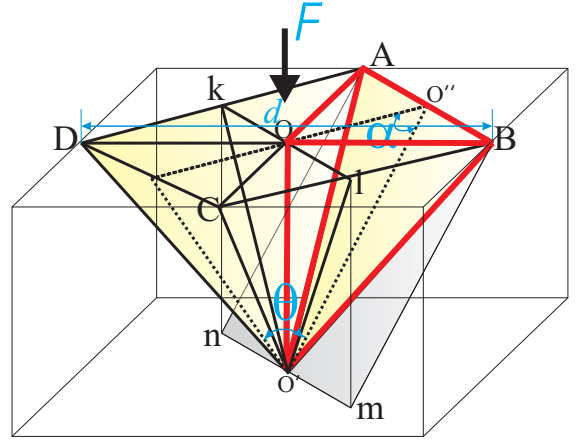}
\end{center}
\caption{(Color online) Illustration of indentation in terms of the
squared diamond pyramid indenter. The red framework highlights one
of four triangular based pyramid indenters. } \label{fig1d}
\end{figure}

\section{Model and Results}
According to Vicker \cite{Gilman}, the hardness of \emph{H}$_v$ is
the ratio between the load force applied to the indenter, \emph{F},
and  the indentation surface area:
\begin{equation} \label{1}
H_v = \frac{2Fsin({\theta/2})}{d^2},
\end{equation}
where \emph{d} and $\theta$ are the mean indentation diagonal and
angle between opposite faces of the diamond squared pyramid
indenter, respectively (Fig. \ref{fig1d}). In order to derive our
model, we first assume that (\emph{i}) the diamond squared pyramid
indenter can be divided into four triangular based pyramid indenters
and that (\emph{ii}) the Vicker's hardness is measured within the
elastic scale. Then, for each triangular based pyramid, one can
define the shear modulus $G$ as,
\begin{equation} \label{2}
G = \frac{F}{4Atan(\alpha)}
\end{equation}
which specifies the ratio between shear stress and the shear stain.
In terms of our model the exact shear area $A$ on which the shear
force (\emph{F}) acts is unknown. But, the deformation area $A^*$
[$A^*$=$\frac{1}{8}d^2tan(\alpha)$] delimited by the klO$^\prime$
triangle is well defined by the indentation geometry. Therefore, we
can express the exact shear area ($A$) as:
\begin{equation} \label{2.1}
A = cA^* = \frac{c}{8}d^2tan(\alpha)
\end{equation}
where \emph{c} is the proportional coefficient. It is clear that
under elastic shear deformation the deformation area (\emph{A}$^*$)
will be extremely small. However, upon real hardness measurements
the deformation area (\emph{A}$^*$) should be large enough so that
the coefficient $c$ can be safely neglected and \emph{A} $\approx$
\emph{A}$^*$. Under this assumption, the equation (\ref{2}) can be
revised as following,

\begin{equation} \label{3}
G = \frac{2F}{d^2tan^2(\alpha)}
\end{equation}

Combining equations (\ref{1}) and (\ref{3}), the Vicker's hardness
reads
\begin{equation} \label{4}
H_v = Gtan^2(\alpha)sin(\theta/2) = 0.92Gtan^2(\alpha)
\end{equation}
where the term sin($\theta$/2) is intrinsically determined by the
indenter itself, which can be considered as a constant (originated
from the Vicker's hardness, see equation (\ref{1})). For the diamond
squared pyramid indenter with $\theta$ = 136$^\circ$ then
sin($\theta$/2) is equal to 0.92 for Vicker's hardness measurement.
In an ideal form of indentation, tan($\alpha$) = 0.404 because of
$\alpha$ = $\frac{(\pi-\theta)}{2.0}$ (c.f., Fig. \ref{fig1d}).
Therefore, the equation (\ref{4}) can be simplified as,

\begin{equation} \label{5}
H_v = 0.151G.
\end{equation}

Equation (\ref{5}) represents a robust theoretical evidence of the
linear correlation behavior observed by Teter\cite{Teter}, as
reflected by the data shown in Fig. \ref{fig1}(b). Residual
discrepancies should be mainly attributed to the neglection of
plastic deformation effects. Remarkably, we  found that Eq.
(\ref{5}) is also valid for BMGs. Using the experimental shear
modulus $G$ = 38.6 (36.6) GPa \cite{7} for
Pd$_{40}$Ni$_{40}$P$_{20}$ BMG, the estimated Vicker's hardness is
5.83 (5.53) GPa, in consistency with the experimental value of 5.38
GPa \cite{7}. Similarly, for
Fe$_{41}$Co$_7$Cr$_{15}$Mo$_{14}$C$_{15}$B$_6$Y$_2$ BMG by using the
experimental $G$ (84.3 GPa \cite{1}) we obtained $H_v$ = 12.7 GPa,
in nice accord with the measured Vicker's hardness of 12.57 $\pm$
0.22 GPa \cite{1}. As illustrated in the inset of Fig.
\ref{fig1}(b), the agreement with the experimental values is highly
satisfactory for all 37 BMGs collected here (see Table \ref{tab:1}).
Considering that BMGs are brittle materials without plastic
fractures the data in the inset of Fig. \ref{fig1}(b) strongly
convey that our proposed formula (Eq. \ref{5}) is intrinsically
connected to the shear modulus of materials if they fail in an
elastic regime.

\begin{figure}[hbt]
\begin{center}
\includegraphics[width=0.40\textwidth]{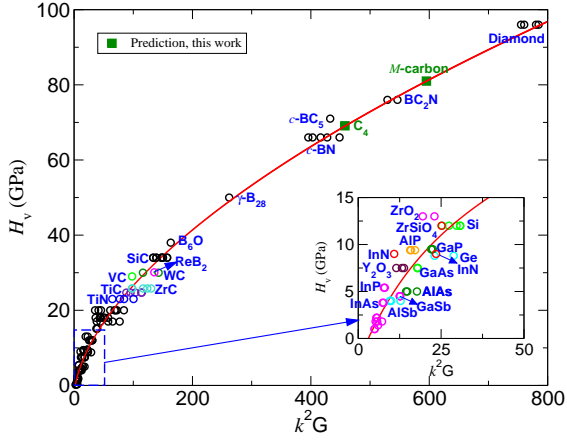}
\end{center}
\caption{(Color online) Experimental Vicker's hardness as a function
of the product $k$$^2$$G$ (\emph{k}=\emph{G}/\emph{B}). All data are
collected from literature (see Tables \ref{tab:2} and \ref{tab:3}).}
\label{fig2}
\end{figure}

\begin{table*}
\centering \caption{The comparison between the predicted Vicker's
hardness by Eq. (\ref{5}) [$H_v$ = 0.151$G$] and the experimental
data for 37 bulk metallic glasses are shown, together with
experimental Young modulus ($E$) and Shear modulus ($G$). The
[XX];[YY] in the first column denotes the reference numbers: XX is
the reference for elastic constants and YY is the reference for the
Vicker's hardness.}
\begin{tabular}{lcccccccccc}
\hline
Compounds &  $E$ &  $G$  & $H_{calc}$ &  $H_{exp}$ \\
\hline
Fe$_{41}$Co$_7$Cr$_{15}$Mo$_{14}$C$_{15}$B$_{6}$Y$_{2}$ \cite{1} & 226 & 84.3  &  12.73&   12.57\\
Ni$_{50}$Nb$_{50}$ \cite{2} &   132 & 48.2 &   7.26  &  8.93 \\
Ni$_{40}$Cu$_5$Ti$_{17}$Zr$_{28}$Al$_{10}$ \cite{3} & 133.9 &  49.7 &   7.50 &   8.45\\
Ni$_{39.8}$Cu$_{5.97}$Ti$_{15.92}$Zr$_{27.86}$Al$_{9.95}$Si$_{0.5}$
\cite{3} & 117 & 43 & 6.49 & 8.13 \\
Ni$_{40}$Cu$_{5}$Ti$_{16.5}$Zr$_{28.5}$Al$_{10}$ \cite{3} & 122 & 45.2 & 6.83 & 7.84 \\
Ni$_{45}$Ti$_{20}$Zr$_{25}$Al$_{10}$ \cite{3} & 114 & 42 & 6.34 &
7.76
\\
Ni$_{40}$Cu$_{6}$Ti$_{16}$Zr$_{28}$Al$_{10}$ \cite{3} & 111 & 40.9 &   6.18 & 7.65 \\
\{Zr$_{41}$Ti$_{14}$Cu$_{12.5}$Ni$_{10}$Be$_{22.5}$\}$_{98}$Y$_{2}$ \cite{4} & 107.6 & 40.3 & 6.09 & 6.76 \\
Zr$_{54}$Al$_{15}$Ni$_{10}$Cu$_{19}$Y$_{2}$ \cite{4} & 92.1 &   33.8 &   5.10 & 6.49 \\
Zr$_{53}$Al$_{14}$Ni$_{10}$Cu$_{19}$Y$_4$ \cite{4} & 86 & 31.5 &   4.76 &   6.44 \\
Zr$_{41}$Ti$_{14}$Cu$_{12.5}$Ni$_{8}$Be$_{22.5}$C$_1$ \cite{4} &  106 & 39.5 &   5.96 &   6.13 \\
Zr$_{46.75}$Ti$_{8.25}$Cu$_{7.5}$Ni$_{10}$Be$_{27.5}$ \cite{2} &   100 & 37.2 & 5.62 &    6.1 \\
Zr$_{48}$Nb$_{8}$Cu$_{14}$Ni$_{12}$Be$_{18}$ \cite{4} & 93.7 &   34.2 &   5.16  &  6.09 \\
Zr$_{34}$Ti$_{15}$Cu$_{10}$Ni$_{11}$Be$_{28}$Y$_2$ \cite{4} & 109.8 &  41 & 6.19  &   6.07 \\
Zr$_{57}$Nb$_{5}$Cu$_{15.4}$Ni$_{12.6}$Al$_{10}$ \cite{2} & 87.3 &   31.9 &   4.82 &   5.9 \\
Zr$_{48}$Nb$_{8}$Cu$_{12}$Fe$_{8}$Be$_{24}$ \cite{4} & 95.7 &   35.2 &   5.32  &   5.85 \\
Zr$_{40}$Ti$_{15}$Cu$_{11}$Ni$_{11}$Be$_{21.5}$Y$_1$Mg$_{0.5}$
\cite{4} & 94.2 & 34.7 & 5.24 & 5.74
\\
Zr$_{41}$Ti$_{14}$Cu$_{12.5}$Ni$_{10}$Be$_{22.5}$ \cite{2};\cite{4} &  101 & 37.4 &   5.65 &   5.97 \\
Zr$_{41}$Ti$_{14}$Cu$_{12.5}$Ni$_{10}$Be$_{22.5}$ \cite{2};\cite{5} & 101 & 37.4 &   5.65 &   5.4 \\
Zr$_{41}$Ti$_{14}$Cu$_{12.5}$Ni$_{10}$Be$_{22.5}$ \cite{2};\cite{6} & 101 & 37.4  &  5.65  &  5.88 \\
Zr$_{41}$Ti$_{14}$Cu$_{12.5}$Ni$_{10}$Be$_{22.5}$ \cite{2}   & 101 & 37.4  &  5.65 &   5.23 \\
Zr$_{65}$Al$_{10}$Ni$_{10}$Cu$_{15}$ \cite{2} &   83 & 30.3 &   4.58 &   5.6 \\
Zr$_{65}$Al$_{10}$Ni$_{10}$Cu$_{15}$ \cite{2} & 83 & 31 & 4.58 &   5.6 \\
Zr$_{57}$Ti$_5$Cu$_{20}$Ni$_{8}$Al$_{10}$ \cite{2} & 82 & 30.1 &   4.55 & 5.4 \\
Cu$_{60}$Hf$_{10}$Zr$_{20}$Ti$_{10}$ \cite{2} & 101 & 36.9 &   5.57 & 7 \\
Cu$_{50}$Zr$_{50}$ \cite{2} & 88.7 & 32.4 & 4.83 & 5.8 \\
Cu$_{50}$Zr$_{50}$ \cite{2} & 85 & 32 & 4.83 & 5.8 \\
Cu$_{50}$Zr$_{45}$Al$_{5}$ \cite{2} & 102 & 33.3 & 5.03 & 5.4 \\
Pd$_{40}$Ni$_{40}$P$_{20}$ \cite{2};\cite{7} &  108 & 38.6 & 5.83 & 5.38 \\
Pd$_{40}$Ni$_{40}$P$_{20}$ \cite{7} & -- & 36.6 & 5.53 & 5.38 \\
Pd$_{40}$Ni$_{40}$P$_{20}$ \cite{2} & 108 &38.6 & 5.83 & 5.3 \\
Pd$_{40}$Ni$_{10}$Cu$_{30}$P$_{20}$ \cite{2} & 98 & 35.1 & 5.30 & 5 \\
Pd$_{77.5}$Si$_{16.5}$Cu$_6$ \cite{2} & 92.9 & 32.9 & 5.25 & 4.5 \\
Pd$_{77.5}$Si$_{16.5}$Cu$_6$ \cite{2} & 96 & 34.8 & 5.25 & 4.5 \\
Pt$_{60}$Ni$_{15}$P$_{25}$ \cite{2} & 96 & 33.8 & 5.10 & 4.1 \\
Mg$_{65}$Cu$_{25}$Tb$_{10}$ \cite{2} & 51.3 & 19.6 & 2.96 & 2.83  \\
Nb$_{60}$Al$_{10}$Fe$_{20}$Co$_{10}$ \cite{2} & 51.2 & 19.4 & 2.93 & 2.2 \\
Ce$_{70}$Al$_{10}$Ni$_{10}$Cu$_{10}$ \cite{2} & 30 & 11.5 & 1.74 & 1.5 \\
Er$_{55}$Al$_{25}$Co$_{20}$ \cite{8} & 70.72 & 27.08 & 4.09 & 5.45 \\
Dy$_{55}$Al$_{25}$Co$_{20}$ \cite{8} &61.36 & 23.52 & 3.55 & 4.7\\
Tb$_{55}$Al$_{25}$Co$_{20}$ \cite{8} & 59.53 & 22.85 & 3.45 & 4.42 \\
Ho$_{55}$Al$_{25}$Co$_{20}$ \cite{8} & 66.64 & 25.42 & 3.84 & 4.14 \\
La$_{55}$Al$_{25}$Co$_{20}$ \cite{8} & 40.9 & 15.42 & 2.33 & 3.48 \\
La$_{55}$Al$_{25}$Cu$_{10}$Ni$_5$Co$_5$ \cite{8} & 41.9 & 15.6 & 2.36 & 3 \\
Pr$_{55}$Al$_{25}$Co$_{20}$ \cite{8} & 45.9 & 17.35 &  2.62 & 2.58 \\
\hline
\end{tabular}
\label{tab:1}
\end{table*}

It is highly difficult to realistically take plastic deformation
into account in our modeling scheme. However, the indentation after
a real hardness measurement shows the permanent plastic effect,
which is, of course, reflected by the ratio
$\frac{l_{OO{^\prime}}}{l_{OO{^\prime}{^\prime}}}$ [namely, equal to
$\tan(\alpha)$] (see Fig. \ref{fig1d}). Note that the depth of the
indentation, $l_{OO{^\prime}}$, is parallel to the direction of
shear deformation. We reasonably assume that its size should be
closely correlated with the shear modulus of $G$, whereas the
expansion wideness of the indentation, $l_{OO{^\prime}{^\prime}}$,
is perpendicular to the direction of the loading force, hence, with
almost little connection to the shear deformation. Therefore, the
expansion wideness seems to reflect the ability to resist to
compression effects, a property that should be related to bulk
modulus, $B$. Accordingly, we proposed the following relation,

\begin{equation} \label{6}
tan(\alpha) \propto G/B
\end{equation}
Finally, combining the equations (\ref{4}) and (\ref{6}), the
hardness can be written as,
\begin{equation} \label{7}
H_v \propto G(G/B)^2
\end{equation}
It is interesting to note that in Eq. (\ref{7}) the ratio of $G/B$
is the famous modulus ratio proposed by Pugh in as early as 1954
\cite{Pugh}. In his pioneer work, Pugh derived that the strain at
fracture can be measured as  $\epsilon \propto (B/G)^2$. Indeed,
hardness can be defined as the resistance to the applied stress at
the critical strain of $\epsilon$ that the system can sustain before
yielding it to fracture. This clearly provides fundamental support
for our model (Eq. \ref{7}). Importantly, Pugh also highlighted a
relation between the elastic and plastic properties of pure
polycrystalline metals and stated that \emph{G}/\emph{B} is closely
correlated to the brittle and ductility of materials \cite{Pugh}:
the higher the value of \emph{G}/\emph{B} is, the more brittle the
materials would be \cite{Pugh}. Otherwise, the materials are
expected to deform in a ductile manner with a low \emph{G}/\emph{B}
value. This relation has been extensively accepted and applied not
only to metal but also to high-strength materials
\cite{Kishida,Jiang,Gaogh}. In principles, the covalent materials
(such as diamond and \emph{c}-BN) have the highest hardness but they
are obviously brittle with a larger Pugh modulus ratio. The strong
covalent bonds indeed create a significant resistance to initialize
the plastic flow to pin the dislocation, resulting in a quite high
hardness. Conversely, ductile materials with a low Pugh's modulus
ratio are characterized by metallic bonding and low hardness. It is
thus highly reasonable to establish a correlation between hardness
and the modulus ratio \emph{G}/\emph{B} in Eq. (\ref{7}). Thus, we
revise further the Eq. (\ref{7}) in the following form,

\begin{equation} \label{equ1}
H_v = Ck^mG^n; (k = G/B)
\end{equation}
where $H_v$, $G$ and $B$ are the hardness (GPa), shear modulus (GPa)
and bulk modulus (GPa), respectively. The parameter $k$ is the
Pugh's modulus ratio, namely, $k$ = $G$/$B$. \emph{C} is a
proportional coefficient. In order to derive the parameters
\emph{C}, \emph{m} and \emph{n}, we first selected ten materials
with diamond-like (diamond, c-BN, $\beta$-SiC, Si and Ge),
zinc-blende (ZrC and AlN) and rock-salt structures (GaP, InSb and
AlSb) because their hardness, bulk and shear moduli are well-known
(see Tables \ref{tab:2} and \ref{tab:3}). By analyzing these data we
found that $C$=1.887, \emph{m}=1.171 and \emph{n} = 0.591. Hence,
Eq. (\ref{equ1}) becomes,

\begin{equation} \label{equ2}
H_v = 1.887k^{1.171}G^{0.591} \approx 1.887 (k^2G)^{0.585}.
\end{equation}

In order to assess the validity of this relation, we plot in Fig.
\ref{fig2} $H_v$ against $k^2G$ for a series of compounds. These
data show a clear and systematic trend with $k^2$\emph{G} and firmly
establish a direct relation between hardness and $k^2G$. By fitting
the data of Fig. \ref{fig2} and revising further Eq. (\ref{equ2}) we
arrive to the final formula:

\begin{equation} \label{equ3}
H_v = 2(k^2G)^{0.585}-3.
\end{equation}
from which we see that the hardness correlates not only with the
shear modulus but also with the bulk modulus. Physically, the bulk
modulus only measures isotropic resistance to volume change under
hydrostatic strain, whereas shear modulus responses to resistance to
anisotropic shear strain. Although the bulk modulus was thought to
be less directly connected with hardness \cite{Gao02}, the Pugh's
modulus ratio $k$ clearly contributes to the Vicker's hardness.
Equation (\ref{equ3}) demonstrates that, if bulk modulus increases,
hardness would decrease as long as the shear modulus remains
unchanged, and vice versa. This behavior can be understood by the
fact that if the Pugh's modulus ratio $G$/$B$ gets smaller with
increasing bulk modulus, the material would become more ductile. Its
hardness can be thus expected to have a lower value. Taking the
example, the compounds TiN and $\beta$-SiC have almost the same
experimental shear moduli \cite{Yao,Kim} of 187.2 GPa and 191.4 GPa,
respectively. However, the experimental bulk modulus of TiN (318.3
GPa)\cite{Yao,Kim} is larger by about 42\% than that of $\beta$-SiC
(224.7 GPa). In terms of our formula, $\beta$-SiC is found to be
harder than TiN, in agreement with the experimental
observations\cite{Gao02,Guo} [Expt (Calc in this work): TiN with
$H_v$ = 23 (20) GPa and $\beta$-SiC with $H_v$ = 34 (33) GPa].

\begin{figure}[hbt]
\begin{center}
\includegraphics[width=0.40\textwidth]{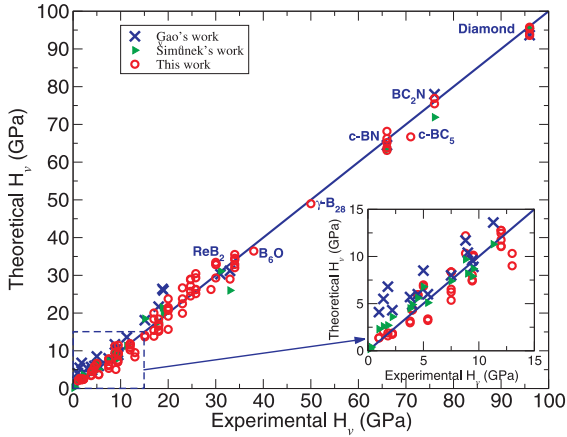}
\end{center}
\caption{(Color online) Correlation between experimental and
theoretical Vickers's hardness ($H_v$) for 39 compounds, as compared
with the estimated data from the models \cite{Gao02,Simunek} (see
Tables \ref{tab:2} and \ref{tab:3}). } \label{fig3}
\end{figure}

\begin{table*}
\centering \caption{Comparison between theoretical values within our
current model and experimental values as compared with available
theoretical findings obtained through the models of Gao \cite{Gao02}
and of Simunek \cite{Simunek}. Furthermore, the bulk and shear
moduli are compiled in this table. In the last column, "$e$" and
"$c$" denote elastic data ($G$ and $B$) from direct experimental and
theoretical calculations, respectively. The Pugh's modulus ratio $k$
is compiled in this table. For details, see text. The [XX];[YY] in
the first column denotes the reference numbers: XX is the reference
for elastic constants and YY is the reference for the Vicker's
hardness.}
\begin{minipage}{18cm}
\begin{tabular}{lcccccccccc}
\hline
Compounds &  G &  B &  k &  $H_{calc}$  & $H_{exp}$ &   $H_{Gao}$  &  $H_{Simunek}$ \\
\hline Diamond \cite{Yao};\cite{Gao02}  & 535.5 &  442.3 &  1.211 &
95.7 &  96 & 93.6 & 95.4 &    e \\
Diamond \cite{Yao};\cite{Gao02}  &548.3 &   465.5 &   1.178 &   93.9
& 96 & 93.6 & 95.4 & c \\
Diamond \cite{12};\cite{Gao02}  & 520.3 &  431.9 &   1.205 &  93.5 &
96 & 93.6 & 95.4 & c \\
Diamond \cite{Teter};\cite{Gao02} &  535.0 &   443.0 &  1.208 & 95.4
&   96 & 93.6 & 95.4 & e \\
BC$_2$N \cite{14};\cite{15} & 446.0 &  403.0  &  1.107 &  76.9 & 76
& 78 &
71.9 & c \\
BC$_2$N \cite{Teter};\cite{15} & 445.0 &  408.0 &  1.091  & 75.4  &
76 &
78 & 71.9 & e \\
BC$_5$ \cite{16};\cite{17} & 394.0 & 376.0 &   1.048 &   66.7  &  71 &   & &      c\\
$c$-BN [18;\cite{Gao02} & 405.4 & 400.0 & 1.014 &  65.2 &   66 &
64.5 &    63.2
& e \\
$c$-BN \cite{Yao};\cite{Gao02} & 403.4 & 403.7 & 0.999 & 63.8 &
66 & 64.5 &    63.2&
c \\
$c$-BN \cite{Yao};\cite{Gao02} & 382.2 & 375.7 & 1.017 & 63.1 & 66 &
64.5 & 63.2 & c
\\
$c$-BN \cite{19};\cite{Gao02} & 404.4 & 384.0 & 1.053 & 68.2 & 66 & 64.5 & 63.2 & c \\
$c$-BN \cite{Teter};\cite{Gao02} & 409.0 & 400.0 & 1.023 & 66.2 & 66 & 64.5 & 63.2 & e \\
$\gamma$-B$_{28}$ \cite{20};\cite{21} & 236.0 & 224.0 & 1.054 & 49.0 & 50 & && c \\
B$_{6}$0 \cite{Teter};\cite{22} & 204.0 & 228.0 &0.895 & 36.4& 38&  & & e \\
$\beta$-SiC \cite{Yao};\cite{Gao02} & 191.4 & 224.7 & 0.852 & 32.8 &
34 & 30.3 & 31.1
& e \\
$\beta$-SiC \cite{Yao};\cite{Gao02} & 196.6 & 224.9 & 0.874 & 34.5 &
34 & 30.3 &31.1
& c \\
$\beta$-SiC \cite{23};\cite{Gao02} & 190.2 & 209.2 &0.909& 35.5& 34 &30.3& 31.1& c\\
$\beta$-SiC \cite{24};\cite{Gao02} & 186.5 & 220.3 & 0.846 & 32.0 &
34 & 30.3 & 31.1
& e \\
$\beta$-SiC \cite{Teter};\cite{Gao02} & 196.0 & 226.0 & 0.867 & 34.1
& 34 & 30.3 &
31.1& e \\
SiO$_2$ \cite{Teter};\cite{Teter} & 220.0 & 305.0 & 0.721 & 29.0 & 33 & 30.4 & & e \\
ReB$_2$ \cite{25};\cite{26} & 273.0 & 382.0 & 0.715 & 32.9 & 30.1 & & & e \\
WC \cite{27};\cite{Gao02} & 301.8 & 438.9 & 0.688 & 33.4 & 30 & 26.4 & 21.5 & e \\
WC [9;\cite{Gao02} & 282.0 & 439.0 & 0.642 & 29.3 & 30 & 26.4 & 21.5 & e \\
B$_4$C \cite{28};\cite{Teter} & 192.0 & 226.0 & 0.850 & 32.8 &
30\footnote{B$_4$C was suggested to be very hard in Ref. \cite{64}.
Mukhanov {\em et al} recently predicted that the Vicker's hardness
of B$_4$C was 45.0 GPa \cite{63} in agreement with the reported
experimental data of 45 GPa in Ref. \cite{65} (see Table 1 in Ref.
\cite{63}). However, we also noted that Teter \cite{Teter} ever
summarized the hardness of B$_4$C with a value of 30$\pm$2 GPa. In
addition, the experimental value of 32--35 GPa was recently
summarized in the handbook \cite{66}. Therefore, here we quoted the
experimental Vicker's hardness of 30 GPa, as summarized by Teter in
Ref. \cite{Teter}.}
& & & e \\
VC [This work];\cite{Simunek} & 209.1 & 305.5 & 0.685 & 26.2 & 29 &
& 27.2 & c
\\
ZrC \cite{29};\cite{30} & 169.7 & 223.1 & 0.761 & 26.3 & 25.8 & & & e \\
ZrC \cite{31};\cite{30} &182.5 & 228.3 & 0.799 & 29.4 & 25.8 & & &c \\
ZrC \cite{31};\cite{30} & 185.9 & 228.0 & 0.815 & 30.5 & 25.8 & & &c \\
ZrC \cite{32};\cite{30} & 169.6 & 223.3 & 0.759 & 26.2 & 25.8 & & & e \\
ZrC \cite{Teter};\cite{30} & 166.0 & 223.0 & 0.744 & 25.2 & 25.8 & & & e \\
TiC \cite{29};\cite{Simunek} & 182.2 & 242.0 & 0.753 & 27.1 & 24.7 & & 18.8 & e \\
TiC \cite{33};\cite{Simunek} & 176.9 & 250.3 & 0.707 & 24.5 & 24.7 &&18.8 & c \\
TiC \cite{34};\cite{Simunek} & 198.3 & 286.0 & 0.693 & 25.8 & 24.7 && 18.8 & c \\
TiC \cite{35};\cite{Simunek} & 187.8 & 241.7 & 0.777 & 28.8 & 24.7 && 18.8 & e \\
TiC \cite{Teter};\cite{Simunek} & 188.0 & 241.0 & 0.780 & 29.0 & 24.7 && 18.8 & e \\
TiN \cite{36};\cite{30} & 183.2 & 282.0 & 0.650 & 22.5 & 23 & & 18.7 & c \\
TiN \cite{37};\cite{30} & 187.2 & 318.3 & 0.588 & 20.0 & 23 & & 18.7 & e \\
TiN \cite{38};\cite{30} & 205.8 & 294.6 & 0.699 & 26.7 & 23 & & 18.7 & c \\
TiN \cite{39};\cite{30} & 207.9 & 326.3 & 0.637 & 23.8 & 23 & & 18.7 & c \\
RuO$_2$ \cite{40};\cite{Gao02} & 142.2 & 251.3 & 0.566 & 15.7 & 20 & 20.6 & & c \\
RuO$_2$ \cite{41};\cite{Gao02} & 173.0 & 248.0 & 0.698 & 23.7 & 20 & 20.6 & & c \\
Al$_2$O$_3$ \cite{19};\cite{Gao02} & 161.0 & 240.0 & 0.671 & 21.5 & 20 & 20.6 & & c \\
Al$_2$O$_3$ \cite{19};\cite{Gao02} & 160.0 & 259.0 & 0.618 & 19.2 & 20 & 20.6 & & c \\
Al$_2$O$_3$ \cite{42};\cite{Gao02} & 164.0 & 254.0 & 0.646 & 20.7 & 20 & 20.6 & &e \\
Al$_2$O$_3$ \cite{Teter};\cite{Gao02} & 162.0 & 246.0 & 0.659 & 21.1 & 20 & 20.6 & &e \\
NbC \cite{43};\cite{Simunek} & 171.0 & 333.0 & 0.513 & 15.5 & 18 & &  18.3 & c \\
NbC \cite{32};\cite{Simunek} & 171.7 & 340.0 & 0.505 & 15.2 & 18 & & 18.3 & e \\
AlN \cite{19};\cite{Gao02} & 134.7 & 206.0 & 0.654 & 18.4 & 18 & 21.7 & 17.6 & c \\
AlN \cite{44};\cite{Gao02} & 130.2 & 212.1 & 0.614 & 16.5 & 18 &
21.7 & 17.6 & c\\
\hline
\end{tabular}
\end{minipage}%
\label{tab:2}
\end{table*}

\begin{table*}
\centering \caption{(Table \ref{tab:2} continued)}
\begin{tabular}{lcccccccccc}
\hline
Compounds &  G &  B &  k &  $H_{calc}$  & $H_{exp}$ &   $H_{Gao}$  &  $H_{Simunek}$ \\
\hline AlN \cite{45};\cite{Gao02} & 123.3 & 207.5 & 0.594 & 15.2 &
18 &
21.7 & 17.6 & c\\
AlN \cite{46};\cite{Gao02} & 132.0 & 211.1 & 0.625 & 17 & 18 & 21.7 & 17.6&  e \\
AlN \cite{Teter};\cite{Gao02} & 128.0 & 203.0 & 0.631 & 16.9 & 18 &
21.7 & 17.6 & e\\
NbN \cite{47};\cite{Simunek} & 155.9 & 292.0 & 0.534 & 15.4 & 17 & & 19.5 & e \\
NbN \cite{Teter};\cite{Simunek} & 156.0 & 315.0 & 0.495 & 13.9 & 17 & & 19.5 &   e \\
 HfN
\cite{Yao};\cite{30} & 186.3 & 315.5 & 0.591 & 20.0 & 17 &&& c \\
HfN \cite{Yao};\cite{30} & 164.8 & 278.7 & 0.591 & 18.4 & 17 &&& c \\
GaN \cite{48};\cite{Gao02} &105.2 & 175.9 & 0.598 & 13.7 & 15.1 & 18.1 & 18.5 & e\\
GaN \cite{Teter};\cite{Gao02} & 120.0 & 210.0 & 0.571 & 14.1 & 15.1 & 18.1 & 18.5 & e\\
ZrO$_2$ \cite{19};\cite{Gao02} & 88.0 & 187.0 & 0.471 & 8.4 & 13 & 10.8& & c \\
ZrO$_2$ \cite{49};\cite{Gao02} & 93.0 & 187.0 & 0.497 & 9.5 & 13 & 10.8 & & e \\
Si \cite{29};\cite{Gao02} &66.6 & 97.9 & 0.680 & 11.8 & 12 & 13.6 & 11.3 & e \\
Si \cite{50};\cite{Gao02} & 64.0 & 97.9 & 0.654 & 10.9 & 12 & 13.6 & 11.3 &   c\\
 Si
\cite{50};\cite{Gao02} & 63.2 & 90.7 & 0.697 & 11.8 & 12 & 13.6 & 11.3 & c \\
Si \cite{51};\cite{Gao02} & 61.7 & 96.3 & 0.640 & 10.2 & 12 & 13.6 & 11.3 & c\\
Si \cite{51};\cite{Gao02} & 61.7 & 89.0 & 0.693 & 11.5 & 12 & 13.6 & 11.3 & c\\
 GaP
[52;\cite{Gao02} & 55.7 & 88.2 & 0.631 & 9.3 & 9.5 & 8.9 & 8.7 & e \\
GaP \cite{29};\cite{Gao02} & 55.8 & 88.8 & 0.628 & 9.2 & 9.5 & 8.9 & 8.7 & e \\
GaP \cite{27};\cite{Gao02} & 56.1 & 88.6 & 0.633 & 9.4 & 9.5 & 8.9 & 8.7 & e \\
GaP \cite{53};\cite{Gao02} & 61.9 & 89.7 & 0.690 & 11.5 & 9.5 & 8.9 & 8.7 & c \\
AlP \cite{54};\cite{Gao02} & 49.0 & 86.0 & 0.570 & 7.1 & 9.4 & 9.6 & 7.9 & e \\
AlP \cite{53};\cite{Gao02} & 51.8 & 90.0 & 0.575 & 7.5 & 9.4 & 9.6 & 7.9 & c \\
AlP \cite{55};\cite{Gao02} & 48.8 & 86.0 & 0.567 & 7.0 & 9.4 & 9.6 & 7.9 & c \\
InN \cite{48};\cite{Simunek} & 55.0 & 123.9 & 0.444 & 5.1 & 9 & 10.4 & 8.2 & c \\
InN \cite{54};\cite{Simunek} & 77.0 & 139.6 & 0.552 & 9.7 & 9 & 10.4 & 8.2 & c \\
Ge \cite{50};\cite{Simunek} & 53.1 & 72.2 & 0.736 & 11.3 & 8.8 & 11.7 & 9.7&  c\\
Ge \cite{50};\cite{Simunek} & 43.8 & 60.3 & 0.726 & 9.5 & 8.8 & 11.7 & 9.7 & c \\
GaAs \cite{27};\cite{Gao02} & 46.5 & 75.0 & 0.620 & 7.8 & 7.5 & 8.0 & 7.4 & e \\
GaAs \cite{52};\cite{Gao02} & 46.7 & 75.5 & 0.619 & 7.8 & 7.5 & 8.0 & 7.4 & e \\
GaAs \cite{29};\cite{Gao02} & 46.7 & 75.4 & 0.619 & 7.8 & 7.5 & 8.0 & 7.4 & e \\
Y$_2$O$_3$ \cite{Yao};\cite{Gao02} & 72.5 & 166.0 & 0.437 &   6.3 & 7.5 & 7.7 & &c \\
Y$_2$O$_3$ \cite{Yao};\cite{Gao02} & 62.7 & 146.5 & 0.428 & 5.3 & 7.5 & 7.7& & c \\
Y$_2$O$_3$ \cite{56};\cite{Gao02} & 66.5 & 149.3 & 0.445 & 6.0 & 7.5 & 7.7 & & e \\
InP \cite{52};\cite{Gao02} & 34.3 & 71.1 & 0.483 & 3.8 & 5.4 & 6.0 & 5.1 & e\\
InP \cite{27};\cite{Gao02} & 34.4 & 72.5 & 0.475 & 3.6 & 5.4 & 6.0 & 5.1 & e \\
AlAs \cite{57};\cite{Gao02} & 44.8 & 77.9 & 0.575 & 6.7 & 5 & 8.5 & 6.8 & e \\
AlAs \cite{52};\cite{Gao02} & 44.6 & 78.3 & 0.569 & 6.5 & 5 & 8.5 & 6.8 & e \\
GaSb \cite{52};\cite{Simunek} & 34.2 & 56.3 & 0.607 & 5.8 & 4.5 & 6.0 & 5.6 & e\\
GaSb \cite{29};\cite{Simunek} & 34.1 & 56.4 & 0.606 & 5.8 & 4.5 &
6.0 & 5.6 & e
\\
GaSb \cite{27};\cite{Simunek} & 34.3 & 56.3 & 0.608 & 5.8 & 4.5 & 6.0 & 5.6 & e\\
AlSb \cite{58};\cite{Gao02} & 31.5&  56.1& 0.561& 4.6 &4 &4.9 &4.9 & c \\
AlSb \cite{52};\cite{Gao02} & 31.9& 58.2 & 0.549& 4.5 &4 &4.9 & 4.9 & e \\
AlSb \cite{29};\cite{Gao02} & 32.5 & 59.3 & 0.548 & 4.6 & 4 & 4.9 & 4.9 & e \\
 AlSb
\cite{27};\cite{Gao02} & 31.9 & 58.2 & 0.549 & 4.5 & 4 & 4.9 & 4.9 & e \\
InAs \cite{52};\cite{Gao02} & 29.5 & 57.9 & 0.509 & 3.6 & 3.8 & 5.7 & 4.5& e \\
 InAs
\cite{29};\cite{Gao02} & 29.5& 59.1& 0.499& 3.4& 3.8& 5.7& 4.5& e\\
 InSb \cite{27};\cite{Gao02} & 23.0&
46.9& 0.490& 2.4& 2.2& 4.3& 3.6& e \\
InSb \cite{52};\cite{Gao02} &22.9 &46.5 &0.492 &2.4 &2.2 &4.3 &3.6& e\\
 InSb \cite{29};\cite{Gao02} &
22.9 & 46.0 & 0.498& 2.5& 2.2& 4.3& 3.6 &e \\
ZnS \cite{29};\cite{Simunek} &32.8 &78.4 &0.418 &2.5 &1.8& &2.7 &e \\
ZnS \cite{27};\cite{Simunek} &31.5 &77.1 &0.408 &2.3 &1.8& &2.7 &e \\
ZnSe \cite{27};\cite{Simunek}& 28.8 &63.1& 0.456& 2.7 &1.4 &&2.6 &e \\
ZnTe \cite{27};\cite{Simunek} &23.4 &51.0 &0.459 &2.1 &1 &&2.3 &e \\
ZnTe \cite{29};\cite{Simunek} &23.4& 51.0 &0.459 &2.1 &1 &&2.3 &e\\
\hline
\end{tabular}
\label{tab:3}
\end{table*}

To further assess the performance of our model (Eq. \ref{equ3}) we
show in Fig. \ref{fig3} a comparison between the estimated and
experimental values for a series of compounds (see Tables
\ref{tab:2} and \ref{tab:3}), confirming a good agreement. Also WC,
which is wrongly found to be a superhard (49 GPa) material within
Teter's linear correlation, is now predicted to have a Vicker's
hardness of 29.3 GPa in very good accordance with experimental value
(30 GPa \cite{Haines}). In particular, Figs. \ref{fig2} and
\ref{fig3} convey that our proposed formula reproduced very well the
Vicker's hardness for all well-known superhard materials (Daimond
\cite{Gao02,Simunek}, BC$_2$N \cite{Teter,Gao02,Simunek,Chang},
$c$-BN \cite{Gao02,Simunek}, $c$-BC$_5$ \cite{17}, and
$\gamma$-B$_{28}$ \cite{21,Oganov}).

The interesting case is the compound of ReB$_2$, which was thought
to be superhard \cite{59}. Although its Vicker's hardness was
debated extensively \cite{59, 60, 61}, there is now a wide-accepted
consensus that its Vickers' hardness of 30.1 GPa at the large
loading force of 4.9 N \cite{25,26}. Using the experimentally
measured bulk and shear moduli \cite{25} ($B$ = 273 GPa and $G$ =
382 GPa) and in terms of our Eq. (\ref{equ3}), the Vicker's hardness
is derived to be 32.9 GPa, in nice agreement with the experimental
data \cite{25}.

Another attention has to be paid to the case of B$_6$O. Using the
experimental bulk and shear moduli ($B$ = 230 GPa and $G$ = 206 GPa)
\cite{62}, its Vicker's hardness is calculated to be 36.7 GPa within
our current model. This value is well within the scale of the
experimentally measured results from 32 to 38 GPa \cite{22,62} for
polycrystalline boron suboxide sintered samples, although a Vicker's
hardness of 45 GPa was reported for the single crystals under a
loading force of 0.98 N \cite{22}. Indeed, the light loading force
of 0.98 N is not large enough to obtain a real hardness. It is thus
expected to have a smaller hardness if a loading force larger than
0.98 N is applied. Our estimated value for B$_6$O is also in good
agreement with the derived value of 37.3 GPa through a very recent
thermodynamic model of hardness \cite{63}.

We further estimated two more phases of carbon (C$_4$ and
$M$-carbon), which were suggested to be superhard
\cite{Mao,Ma,Chen,Umemoto,Xu}. Utilizing elastic shear and bulk
moduli obtained in Ref. \cite{Chen}, the Vicker's hardness of C$_4$
is calculated to be 69.0 GPa ({\em c.f.}, Fig. \ref{fig2}) that is
comparable to the superhard $c$-BN. Moreover, through using the
calculated bulk and shear moduli ($B$ = 415 GPa and $G$ = 468 GPa
\cite{Zhouxf}) for the $M$-carbon phase, we obtained its Vicker's
hardness of 81.0 GPa ({\em c.f.}, Fig. \ref{fig2}), placing
\emph{M}-carbon in between BC$_2$N and diamond, agreeing well with
the value (83.1 GPa) obtained by \v{S}imunek's model \cite{Ma}.

In addition, from Fig. \ref{fig3} all estimated data are in good
agreement with those obtained from pervious models
\cite{Gao02,Simunek}. Nevertheless, we would like to emphasize that,
although our proposed model can reproduce well the results obtained
by Cao's \cite{Gao02} and \v{S}imunek's \cite{Simunek} models, the
underlying mechanism is substantially different. Gao's and
\v{S}imunek's models are based on bond properties such as
bond-length, charge density, ionicity and their strengths and
coordinations in crystalline materials. Differently, our model
depends totally on the so-called polycrystalline moduli (bulk and
shear modulus as well as Pugh's modulus ratio), which indeed
response directly to the abilities of resistance under loading
forces for polycrystalline materials. As demonstrated above, for
polycrytalline materials the introduced Pugh's modulus ratio in our
model plays a crucial role in elucidating plastic deformation, which
is intrinsically different from all known semi-empirical hardness
models \cite{Gao02,Simunek,Mukhanov,Li08,Smedskjaer}.

\section{Discussion and Remarks}

The hardness of a material is the intrinsic resistance to
deformation when a force is applied \cite{Tse}. Currently, a formal
theoretical definition of hardness is still a challenge for
materials scientists. The need for alternative superhard and
ultrahard materials for modern technology has brought a surge of
interest on modeling and predicting the hardness of real materials.
In particular, in recent years several different semi-empirical
models for hardness of polycrystalline covalent and ionic materials
have been proposed. Gao's model is mainly based on bond length,
charge density, and ionicity \cite{Gao02}. Simunek's model employs
the strength of the chemical bonds and its framework in crystalline
materials \cite{Simunek}. Mukhanov's model utilizes the
thermodynamical concept of energy density per chemical bond
\cite{Mukhanov}. Li's model is mainly based on the bond
electron-holding energy hardness through electronegativity
\cite{Li08}. Smedskjaer's model correctly predicts the hardness of
multicomponent BMGs through temperature-dependent constraint theory
\cite{Smedskjaer}. It has been noted that all the above models
\cite{Gao02,Simunek,Mukhanov,Li08,Smedskjaer} have two major
limitations: (i) Each one provides a satisfactory description only
for a specific type of materials: Smedskjaer's model treats BMGs
with high degree of structural and topological disorder, whereas all
other methods are essentially applicable to crystalline materials
only, and (ii) they depends on different theoretical assumptions.
Since these methods can only be used to predict the hardness of some
specific materials a unified and general theory capable to account
for the hardness of any material is still missing. If we look back
the history of hardness, it can be easily found in literature that
many scientists (Gilman, Cohen, Pugh and Teter)
\cite{Gilman,Gilman2,Cohen,Pugh,Teter} have tried to create a
correlation between the hardness and elasticity (a well defined
quantity) since 1950s. The most successful empirical correlation was
proposed by Teter in 1998 \cite{Teter}, who suggested that the
hardness shows a quasi-linear correlation with the shear modulus.
However, all these empirical correlations between hardness and bulk
modulus (or shear modulus) turned out to be not fully successful.
The main reason is that hardness indeed is a characteristic of a
permanent plastic deformation, whereas the elasticity corresponds to
the reversible elastic deformation. Therefore, there seems to be a
general consensus on the fact that the hardness in general does not
depends neither on the bulk modulus nor on the shear modulus for
polycrystal materials. Indeed, these correlations were heavily
debated in recent years.

In 1954 Pugh has proposed a relation between the elastic and plastic
properties of pure polycrystalline metals and stated that the Pugh's
modulus ratio ($k$ = \emph{G}/\emph{B}) represents a good criterion
to identify the brittleness and ductility of materials \cite{Pugh}.
It was found that material with a large \emph{k} behaves in a more
brittle manner, and that the higher the value of \emph{k} is, the
more brittle the materials are \cite{Pugh}. On the other side,
materials with a low \emph{k} are expected to deform in a more
ductile way. Basically, as evidenced from Tables \ref{tab:2} and
\ref{tab:3} we found that the Pugh's modulus ratio, \emph{k}, can be
correlated with hardness. The hardest material, diamond, has the
highest \emph{k} value of about 1.2 and all widely accepted
superhard materials have a highly large \emph{k} value larger than
1.0. In addition, from Tables \ref{tab:2} and \ref{tab:3} one can
see that with the progressive decrease of the Vicker's hardness the
\emph{k} values get progressively smaller. Unlike the moduli of $G$
and $B$, which only measures the elastic response, the Pugh's
modulus ratio seems to correlate much more reliably with hardness
because it responses to both elasticity and plasticity, which are
the most intrinsic features of hardness.

Therefore, through the introduction of the classic Pugh modulus
ratio proposed in 1954 \cite{Pugh} and following the spirit of
Teter's empirical correlation, we have constructed a theoretical
model of hardness. We proposed a new formula to calculate Vicker's
hardness, \begin{math} H_v = 2(k^2G)^{0.585}-3\end{math}, for
polycrytalline materials. The most important aspect of our formula
is that it correctly predict the hardness of all compounds dataset
considered in several recent models
\cite{Gilman2,Gao02,Simunek,Mukhanov,Li08,Teter,Gilman} just under
the condition of knowing the corresponding bulk and shear moduli.
Furthermore, we proposed that, if a material is intrinsically
brittle (such as BMGs), its Vicker's hardness linearly correlates
with the shear modulus ($H_v$ = 0.151 $G$). On the one hand, this
simplified form provides a robust theoretical evidence why the
famous empirical quasi-linear correlation observed by Teter in 1998
are right for some materials. On the other hand, we found for the
first time the Vicker's hardness can be linearly correlated with the
shear modulus for BMGs (see Table \ref{tab:1}). This is somehow
unexpected because for BMGs there exists a universal correlation
between the Yough's modulus and the Vicker's hardness as documented
in Refs. \cite{2,Whang,Ohtsuki}.

Finally, we still want to point out that our model (Eq. \ref{equ3})
may be not accurate to predict the hardness of pure metals (or
metallic-bonding dominated materials with a highly low Pugh's
modulus ratio). For instance, the hardness of fcc Al is estimated to
be 1.3 GPa, which is significantly larger than the experimental
Vicker's hardness of 0.167 GPa. This is mainly due to the fact that
ductile metals can locally accumulate plastic deformation prior to
fracture, which has not been considered in our model.

In summary, via the aid of Pugh's modulus ratio, our work provides
the firm evidence that the hardness not only correlates with shear
modulus as observed by Teter \cite{Teter}, but also with bulk
modulus as observed by Gilman and Cohen \cite{Cohen,Gilman,Gilman2}.
By retaining the fundamental aspects of the previous proposed
models, our model clearly demonstrates that the hardness of BMGs is
intrinsically based on the same fundamental basis as the crystalline
materials. Given the fact the elastic bulk and shear moduli can be
accurately calculated by the state-of-the-art first-principles
calculations, we believe that our finding is important for the
community to design and develop ultrahard/superhard materials and
high-performance high-strength structural materials.

{\bf Acknowledgements} We greatly appreciate useful discussions with
Profs. Zhang Zhefeng and Luo Xinghong in the IMR. X. -Q. C.
acknowledges the support from the ``Hundred Talents Project'' of CAS
and the NSFC (Grant No. 51074151). The authors also acknowledge the
computational resources from the Supercomputing Center (including
its Shenyang Branch in IMR) of Chinese Academy of Sciences and the
local HPC cluster of the Materials Process Modeling Division in the
IMR.

\end{document}